\documentclass[useAMS,usenatbib,usegraphicx]{mn2e}

\usepackage{amssymb}
\usepackage{fixltx2e}
\usepackage{lscape}
\usepackage{rotating}

\newcommand{\cov}{\mathop{\rm Cov}\nolimits}

\newcommand{\FAP}{{\rm FAP}}

\title[Benchmarking amateur TTV exoplanets detection]{Benchmarking the power
of amateur observatories for TTV exoplanets detection}

\author[Baluev et~al.]{%
Roman V. Baluev$^{1,2}$\thanks{E-mail: r.baluev@spbu.ru}, Evgenii N. Sokov$^1$, Vakhit Sh. Shaidulin$^{2,1}$, Iraida A. Sokova$^1$,
\newauthor
Hugh R. A. Jones$^3$, Mikko Tuomi$^{3,4}$, Guillem Anglada-Escud\'e$^{3,5}$, Paul Benni$^6$,
\newauthor
Carlos A. Colazo$^7$, Matias E. Schneiter$^8$, Carolina S. Villarreal D'Angelo$^8$,
\newauthor
Artem Yu. Burdanov$^9$, Eduardo Fern\'andez-Laj\'us$^{10,11}$\thanks{
Visiting Astronomer, Complejo Astron\'omico El Leoncito operated under agreement
between the Consejo Nacional de Investigaciones Cient\'ificas y  T\'ecnicas de
la Rep\'ublica Argentina and the National Universities of La
Plata, C\'ordoba and San Juan.}, \"Ozg\"ur Ba\c{s}t\"urk$^{12}$,
\newauthor
Veli-Pekka Hentunen$^{13}$, and Stan Shadick$^{14}$\\
$^1$Central Astronomical Observatory at Pulkovo of Russian Academy of Sciences, Pulkovskoje
shosse 65, St Petersburg 196140, Russia\\
$^2$Sobolev Astronomical Institute, St Petersburg State University, Universitetskij prospekt
28, Petrodvorets, St Petersburg 198504, Russia\\
$^3$University of Hertfordshire, Centre for Astrophysics Research, Science and
Technology Research Institute, College Lane,\\ Hatfield AL10 9AB, UK\\
$^4$University of Turku, Tuorla Observatory, Department of Physics and
Astronomy, V\"ais\"al\"antie 20, FI-21500, Piikki\"o, Finland\\
$^5$School of Physics and Astronomy, Queen Mary University of London, 327 Mile End Road,
London E1 4NS, UK\\
$^6$Acton Sky Portal (Private Observatory), Acton, MA, USA\\
$^7$Observatorio Astron\'omico, Universidad Nacional de C\'ordoba, Laprida 854, C\'ordoba X5000BGR, Argentina\\
$^8$Instituto de Astronom\'ia Teor\'ica y Experimental, Universidad Nacional de C\'ordoba, Laprida 854, C\'ordoba X5000BGR, Argentina\\
$^9$Kourovka Astronomical Observatory of Ural Federal University, Mira str. 19, Ekaterinburg 620002, Russia\\
$^{10}$Facultad de Ciencias Astron\'omicas y Geof\'isicas - Universidad Nacional de La Plata, Paseo del Bosque S/N - 1900 La Plata,\\ Argentina\\
$^{11}$Instituto de Astrof\'isica de La Plata (CCT La Plata - CONICET/UNLP), Argentina\\
$^{12}$Ankara University, Faculty of Science, Department of Astronomy and Space Science, TR-06100, Tandogan, Ankara, Turkey\\
$^{13}$Taurus Hill Observatory, Warkauden Kassiopeia ry., H\"ark\"am\"aentie 88, 79480 Kangaslampi, Finland\\
$^{14}$Physics and Engineering Physics Department, University of Saskatchewan, 116 Science Place, Saskatoon, Saskatchewan,\\ S7N 5E2, Canada}

\begin{document}

\date{Accepted 2015 April 8.
      Received 2015 April 7;
      in original form 2015 January 27}

\pagerange{\pageref{firstpage}--\pageref{lastpage}} \pubyear{2015}

\maketitle

\label{firstpage}

\begin{abstract}
We perform an analysis of $\sim 80000$ photometric measurements for the following $10$
stars hosting transiting planets: WASP-2, -4, -5, -52, Kelt-1,
CoRoT-2, XO-2, TrES-1, HD~189733, GJ~436. Our analysis includes mainly transit lightcurves
from the Exoplanet Transit Database, public photometry from the literature, and some
proprietary photometry privately supplied by other authors. Half of these lightcurves were
obtained by amateurs. From this photometry we derive $306$ transit timing measurements, as
well as improved planetary transit parameters.

Additionally, for $6$ of these $10$ stars we present a set of radial velocity measurements
obtained from the spectra stored in the HARPS, HARPS-N, and SOPHIE
archives using the HARPS--TERRA pipeline.

Our analysis of these TTV and RV data did not reveal significant hints of additional
orbiting bodies in almost all of the cases. In the WASP-4 case, we found hints
of marginally significant TTV signals having amplitude $10-20$~sec, although their
parameters are model-dependent and uncertain, while radial velocities did not reveal
statistically significant Doppler signals.
\end{abstract}


\section{Introduction}
The first extrasolar planet, orbiting a solar-type star 51~Pegasi, was discovered by
\citet{MayorQueloz95}, based on the precision Doppler measurements of the ELODIE
spectrograph. After that, the number of the detected exoplanets grew continuously,
exceeding $1000$ so far. In fact, 20 years ago a new rapidly-growing domain of fundamental
science was created, devoted to the exoplanet research.

Currently, most of the known exoplanetary candidates were detected by one of two major
techniques: the radial velocity (RV) method or transit method that increased its output in
recent time thanks to the launch of specialized spacecraft CoRoT (ESA) and Kepler (NASA).

Unfortunately,
both the Doppler and transit exoplanet detection methods require sophisticated and
expensive instrumentation and remain practically inaccessible to the majority of the
community. The RV method requires the use of extremely stable spectrographs. There is only
about a dozen of such instruments in the world that are capable of detecting an exoplanet.
The transit detection is also difficult. It requires precise alignment
of the planetary orbit with the observer, and the transit event is rather short in time,
although periodic. To detect a transiting planet, we have to observe lots of stars, and we
necessarily deal with large numbers of null detections. This requires the use of
specialized ground-based robotic telescopes or telescope networks,
or space observatories like the above-mentioned CoRoT and Kepler. Organizing
such a campaign is a difficult task even for professional scientific teams.

In terms of the photometric accuracy and quality, ground based telescopes are no match
to spacecrafts like Kepler. But all space projects have a common disadvantage:
they are severely limited in time, while the sensitivity to weak planetary signals
in the data degrades quickly when the planetary orbital period exceeds the observational
time base. This condition means that Kepler data cannot reliably
detect long-period planets, analogous to the giant planets of Solar System. In general, the
responsibility for follow-up observations always returns to ground-based
observatories. Ground-based observations of planetary transits are much less demanding than
precision radial velocity measurements, and of course much cheaper than projects
like Kepler. In fact, such observations are possible with commercially available equipment
typically used by a significant community of ``amateurs''. Thus, the exoplanetary hunt can
potentially become ``citizen science'' that can trigger a qualitative leap in the field.

However, amateurs are definitely not equipped to undertake classic transit surveys
like e.g. SuperWASP or others. But nonetheless they can provide a useful
scientific contribution by means of the transit timing variation (TTV)
method \citep{HolmanMurray05,Agol05}. In this approach, there is a list of
well defined targets that are continuously monitored. Each target is a known host of a
transiting planet, and its transits are regularly observed. If more planets orbit the host,
they should induce perturbational effects on the motion of the
transiter, causing observable delays to its transits, in comparison with a
strictly periodic ephemeris. The TTV method allows the observation time to be used
more efficiently, because we know what targets should be observed, and when.

This work represents an attempt to determine the practical efficiency of such an approach,
based on the photometry data, taken mainly from the Exoplanet Transit Database (ETD) of the
Czech Astronomical Union, \texttt{http://var2.astro.cz/ETD/}.

\begin{figure*}
\includegraphics[width=\linewidth]{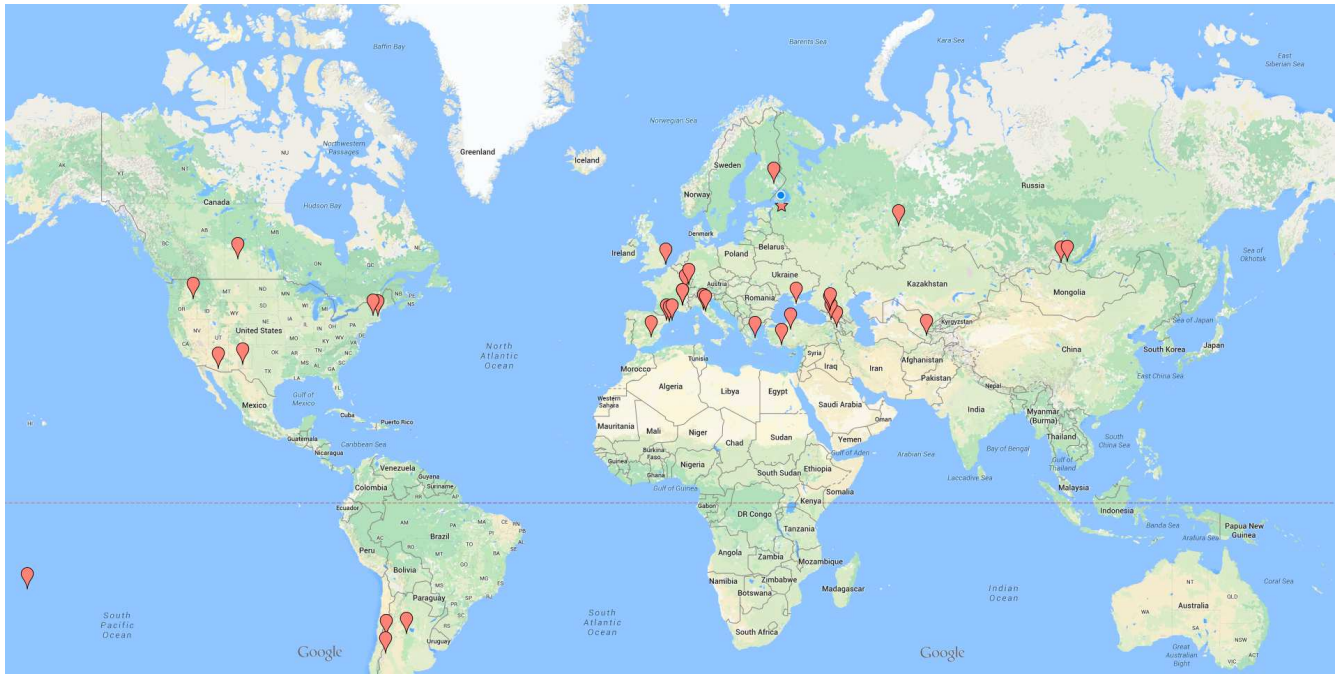}
\caption{World distribution of the observatories regularly contributing to the ETD database.}
\label{fig_map}
\end{figure*}

Currently, about $30$ observatories are regularly contributing
to the database, including amateur as well as professional ones. This network offers
telescopes of different apertures -- from $20$~cm to $2.6$~m. Their locations are shown in
Fig.~\ref{fig_map}. We must note that in terms of technical characteristics of the
telescopes, there is no sharp boundary between the amateur and professional equipment, and
the quality of the observations is often determined by the local astroclimate, which can
have even more important effect than the telescope size. The observational programme
involves currently $\sim 20$ transiter hosts that are more or less regularly observed. For
some of the stars, several years of observations are already available. However, the
transit fitting algorithm used by ETD is criticized for its simplicity and imperfections
\citep[e.g.][]{Petrucci13}. In this work we present a new data reduction pipeline that
handles subtle photometric effects (like the red noise) more accurately. This pipeline was
developed to deal with the photometry of a relatively poor or moderate quality, which is
typical for the amateur data in ETD.

We do not limit ourselves to only amateur observations or only ETD. We also use photometric
data published in the literature. Moreover, for our targets we revealed a moderate
amount of the spectra stored in the HARPS, HARPS-N, and SOPHIE archives, and we derive
RV data from them to provide and independent ``calibration'' of our TTV results.
However, in this work we have no goal to perform a fully self consistent
TTV+RV analysis. Instead, we aim to characterize the accuracy and reliability of the TTV
data that we can derive from just the photometry.

The structure of the paper is as follows. In Sect.~\ref{sec_src} we provide a detailed
description of all the data that we include in our analysis. In Sect.~\ref{sec_analysis} we
introduce the algorithms used to process the photometric data. In Sect.~\ref{sec_ttv} we
present the TTV data derived from the photometry and describe the results of their
periodogram analysis. In Sect.~\ref{sec_trpar} we give the remaining fitted parameters
of the planets considered in the work. In Sect.~\ref{sec_RV} we describe the RV data
obtained for some of our targets on the base of the spectra found in the
HARPS/HARPS-N/SOPHIE archives. In Sect.~\ref{sec_WASP4} we discuss in detail the case of
WASP-4, for which we detected possible hints of a weak TTV signal.

\section{The source data}
\label{sec_src}
Our primary source of the photometric data is the Exoplanet Transit Database (ETD). We
extracted a set of the best transit lightcurves from ETD for $10$ selected stars. Also, we
added in our analysis several public transit lightcurves that we found
in the literature. Thus, our photometric data cover $10$ targets
with $306$ transit lightcurves, and about $80000$ individual photometric measurements. The
public data were presented in the works listed in Table~\ref{tab_ref}, and most are stored
in the Vizier database. Among these $306$ transit lightcurves, $161$ (or roughly half) were
obtained by amateurs, while $65$ are from professional observatories contributed to ETD,
and $80$ were published in the listed literature.

\begin{table}
\caption{Sources of the photometric data (except for ETD).}
\label{tab_ref}
\begin{tabular}{cp{60mm}}
\hline
target & references \\
\hline
WASP-2    & \citealt{Southworth10} \\
WASP-4    & \citealt{Wilson08,Gillon09,Winn09,Southworth09b,Sanchis-Ojeda11,Nikolov12,Petrucci13} \\
WASP-5    & \citealt{Southworth09a} \\
WASP-52   & No \\
Kelt-1    & \citealt{Siverd12} \\
CoRoT-2   & \citealt{Gillon10} \\
XO-2      & \citealt{Fernandez09} \\
TrES-1    & \citealt{Winn07b} \\
HD~189733 & \citealt{Bakos06,Winn07a,Pont07} \\
GJ~436    & \citealt{Bean08b,Shporer09,Stevenson12} \\
\hline
\end{tabular}\\
\end{table}

We need to note that the data from \citep{Sanchis-Ojeda11}
also contain reprocessed photometry from \citep{Winn09}, so the original
\citet{Winn09} data were not actually included in our analysis. Also, we eventually decided
to drop the HST data by \citet{Bean08b}, because they all appeared to cover only small
parts of a transit and could not produce good precision in the derived
mid-times. The photometry presented by \citet{Petrucci13} is not public, but it was kindly
released to us by the authors. The WASP-4 data from \citep{Southworth09b} appeared not very
reliable due to the clock errors of the Danish telescope \citep{Nikolov12,Petrucci13}, and
we believe that the data from \citep{Southworth09a,Southworth10} should be treated with the
same care. However this photometry is rather accurate in itself, and thus it
is still useful in constraining all transit parameters except for mid-times (e.g. by adding
a fittable time offset to these light curves model relative to the other lightcurves). All
timings in the photometric data were transformed to the ${\rm BJD}_{\rm TDB}$ time
stamps by means of the public IDL software developed by
\citet{Eastman10}.

Additionally, we used the precision radial velocity (RV) obtained from the spectra
of the HARPS archive, available for the following targets from our
photometry sample: WASP-2, -4, -5, HD~189733, Corot-2. These spectra were processed with
the advanced HARPS--TERRA pipeline that offers an improved RV accuracy
\citep{AngladaEscudeButler12}. The quality of these RV data, and their number per a target,
appeared not very high. The meaning of the columns in the attached RV data file is given in
Table~\ref{tab_RV}.

\begin{table}
\caption{Explanation of the RV data file}
\label{tab_RV}
\begin{tabular}{cp{7cm}}
\hline
column & explanation   \\
\hline
1      & observation time (BJD) \\
2, 3   & derived radial velocity and its uncertainty \\
4      & a standardized name of the input RV file that includes the target name and the name of the spectrograph \\
\hline
\end{tabular}
\end{table}

For GJ~436 we found a large amount of high-quality HARPS RV data, similar to the one
considered by \citet{Lanotte14}, as well as a large Keck RV time series recently published
by \citet{Knutson14}. We acknowledge that cases like this deserve to be investigated
detailedly in a separate work. In this paper we only present our TTV and TERRA RV data for
GJ436, processed in the common simplistic way as for the other stars.

In this work we do not perform the joint transits+RV fits, because it appeared that to
perform such an analysis at a desirable level of quality, we must provide a solution
to several non-trivial issues that fall outside the scope of this paper. These
issues include, e.g., the treatment of the correlational structure of the RV
noise, and treatment of the RV points affected by the Rossiter-McLaughlin effect. In
this work we used the RV data mainly as an independent source of information.

\section{Methods of the transit lightcurves analysis}
\label{sec_analysis}
\subsection{Details of the transit model and its parametrization}
\label{sec_model}
Let us first adopt the approximation that the transiting planet moves along a straight line
with constant velocity, thus neglecting the orbital curvature of its actual trajectory
during the transit. The model of the linear motion easily predicts the
separation between the centers of planet and star disks, $\delta$, as a function
of time. We have $4$ kinematic characteristics of the transit that must be fitted: (i) the
mid-time of the transit $t_c$, (ii) the duration of the transit $t_d$, defined as the time
spent between the first and fourth contacts and (iii) the impact parameter
$b$, measuring the smallest projected separation $\delta$, and (iv) the projected planet
radius $r$ that simultaneously determines the transit depth and the geometry
of the ingress/egress phases. Given these parameters and easy geometric constructions, the
projected separation $\delta$ can be expressed as:
\begin{equation}
\delta(t) = \sqrt{b^2 + \left[(1+r)^2-b^2\right] \tau^2}, \quad \tau = 2 \frac{t-t_c}{t_d}, \quad |b| \leq 1+r.
\label{delta}
\end{equation}
This assumes that the star radius is unity. To simplify the formula for $\delta$ and
simultaneously get rid of the non-trivial definition domain for $b$, our algorithm
adopts internally a replacement of the impact parameter $b$:
\begin{equation}
b = \frac{p}{\sqrt{1+p^2}} (1+r).
\end{equation}
With the new parameter $p$ we do not need to worry about its domain: any real value of $p$
is physically meaningful. Thus we eliminate the danger that this parameter can walk to a
forbidden domain during the fit. With $p$ replacing $b$, we have
\begin{equation}
\delta(t) = (1+r) \sqrt{\frac{\tau^2+p^2}{1+p^2}}.
\label{delta2}
\end{equation}

However, this approximation of $\delta(t)$ might be inaccurate due to the curvature of the
planet trajectory. In this work we include a correction to this formula, assuming that the
planet moves along a circular orbit. In this approximation, the projected distance can
be expressed by the same formulae~(\ref{delta}) and~(\ref{delta2}), substituting the
following quantity $\tau'$ instead of $\tau$:
\begin{equation}
\tau' = \frac{\sin \tau \alpha}{\sin\alpha}, \qquad \alpha=\pi\frac{t_d}{P} \ll 1,
\label{taucurv}
\end{equation}
where $P$ is the transiter's orbital period. The auxiliary angle $\alpha$ reflects the
curvature of the circular orbit. In fact, this correction induced only a well negligible
effect on our TTV measurements, but nonetheless our results correspond to the model with
this correction included.

The trajectory curvature also depends on the orbital eccentricity, but usually
this eccentricity is difficult to determine from transit observations reliably, and we have
no other option except to assume that it is zero. A zero eccentricity is a
good prior assumption for most of these short-period planets. In the cases when radial
velocity data are also available, the accurate eccentricity information can be
obtained from the joint transit+RV fits, but we do not perform fits of such type in this
work.

After the function $\delta(t)$ is calculated, we approximate the relative flux reduction
with the use of the stellar limb darkening model
by \citet{AbubGost13}. These authors provided theoretical formulae for various types of the
limb darkening effect, as well as a software library written in C. The library
provides subroutines to compute the observed light flux reduction $\Delta L$ as a function
of the eclipsing planet radius $r$ and of the projected separation $\delta$ (assuming that
the star radius is unit), as well as of the limb darkening coefficients that depend on the
selected model. Besides, this library provides partial derivatives of $\Delta L$
with regard to its arguments. These derivatives are necessary to compute the gradient of
the likelihood function, which is also used by the transit fitter.

Thus for a transit lightcurve we have $4$ fittable parameters of the planet or planetary
orbit: the planet/star radii ratio $r$, the transit mid-time $t_c$, the transit duration
$t_d$, and the impact parameter replacer $p$. As the orbital period $P$ is known for all
our planets with a very good accuracy, we treat it as a fixed parameter in the
formula~(\ref{taucurv}).

Additionally, there are two parameters determining the limb darkening model. In this
work we use a quadratic two-term model of the stellar limb darkening with two coefficients
to be determined, $A$ and $B$. The brightness of a point
on the visible stellar disc, observed at a given separation from its center, $\rho$,
is modelled as
\begin{eqnarray}
I(\rho) &=& 1 - A (1-\mu) - B (1-\mu)^2 = \nonumber\\
        &=& 1-A-2B + (A+2B)\mu + B\rho^2, \nonumber\\
        & & \mu=\sqrt{1-\rho^2}.
\label{bright}
\end{eqnarray}

However, in practice the brightness model~(\ref{bright}) can easily turn non-physical,
if the coefficients $A$ and $B$ are allowed to attain arbitrary values. This becomes a
significant problem if our data are polluted by some systematic errors that are always
difficult to foresee in advance. There are a couple of natural basic constraints that we
place on the coefficients $A$ and $B$ to keep the model~(\ref{bright})
physically reasonable:
\begin{equation}
I(\rho) \geq 0, \quad \frac{dI}{d\rho} = \left[ 2B\mu - (A+2B) \right] \frac{\rho}{\mu} \leq 0, \qquad \forall \rho \in [0,1].
\end{equation}
If the second condition (the one on the derivative) is satisfied, $I(\rho)$ is a monotonic
non-increasing function, so the condition $I(\rho) \geq 0$ for $\rho\in [0,1]$ is then
equivalent to $I(1) = 1-A-B \geq 0$. In the second condition, the expression in parenthesis
is a linear function of $\mu$, so to have it always non-positive for $\mu\in [0,1]$, it is
necessary and sufficient to have it non-positive at the boundaries $\mu=0$ (disk limb) and
$\mu=1$ (disk center). Finally, we have total of three elementary inequality constraints on
$A$ and $B$ to satisfy:
\begin{equation}
A+B\leq 1, \quad A+2B\geq 0, \quad A \geq 0.
\label{ldcon}
\end{equation}
Note that the coefficient $B$ is allowed to be negative here (but $B\geq -1$). We do not
put an extra condition $B\geq 0$. Negative values of $B$ mean that the
limb-darkening gradient is diminishing (in absolute value) closer to the limb, although it
never turns positive (any ``limb brightening'' is disallowed).

We may satisfy~(\ref{ldcon}) by means of making a smooth replacement of the parameters $A$
and $B$, such that the formulae of the replacement would
disallow the conditions~(\ref{ldcon}) to be broken. This can be reached, for example, by
the following trigonometric replacement:
\begin{equation}
A = \sin^2\theta\, (1-\cos\varphi), \quad B = \sin^2\theta \cos\varphi.
\end{equation}
Naturally, whatever real values the new auxiliary parameters
$\theta$ and $\varphi$ might attain, the resulting values of $A$ and $B$
always satisfy~(\ref{ldcon}). From the other side, each point $(A,B)$ in the
domain~(\ref{ldcon}) maps to some pair $(\theta,\varphi)$ with real values
of the parameters:
\begin{equation}
\sin^2\theta = A+B, \quad \cos\varphi = \frac{B}{A+B}.
\end{equation}
Note that from~(\ref{ldcon}) it follows that $A+B \geq |B|$, meaning that $A+B$ is never
negative.

Therefore, treating the auxiliary angles $\theta$ and $\varphi$ as primary
fittable parameters, we can satisfy the conditions~(\ref{ldcon}) automatically. The result
of the fitting would correspond to either an internal point of the domain~(\ref{ldcon}), or
to some point on its boundary, if the actual data suggest to move the solution
to a non-physical domain due to e.g. their poor statistical quality or
systematic errors. The boundary points correspond to certain special values of $\theta$ and
$\varphi$ that can be easily identified: $\theta=\pm\pi/2+\pi k$ on the first boundary
of~(\ref{ldcon}), $\varphi=\pi + 2\pi k$ on the second boundary, and $\varphi=2\pi k$ at
the third boundary.

\subsection{Details of the fitting procedure}
\label{subsec_an_g}
In our analysis we are interested in obtaining the TTV deviations with maximum
accuracy achievable with the available photometric data. However, the ETD data, even after
the pre-selection, often have only a moderate quality, performing a complete and
independent fit for each transit lightcurve is not a good option. The accuracy of thus
obtained transit parameters would often be poorly constrained, and this would impact the
accuracy of the fitted TTV data as well. Moreover, the transit fits for individual
lightcurves sometimes do not even converge to a reasonable result, e.g., due to incomplete
coverage of the transit or presence of significant curved trends.

To overcome these issues, we adopt in this work the following approach. We perform a
\emph{joint} fit of all transit lightcurves, available for a given star, assuming that most
of the transit parameters, except for the mid-times, are equal for different
lightcurves. Such a ``shared'' transit parameter is still fittable, taking into account the
constraint of its values being equal between different
lightcurves. The mid-times are fitted individually for each lightcurve, i.e. they remain
unconstrained.

Such an approach uses the full statistical power
of all photometric data, available for a given star, to fit the shape of the transit curve,
while still fitting the TTV offsets of individual transits separately from each other.
We can note the following potential weaknesses of this approach:
\begin{enumerate}
\item It is not taken into account that the limb darkening coefficients depend
on the photometry band, which are different for different lightcurves.
Therefore, the results of such fitting method would refer to some averaged limb darkening,
possibly introducing minor modelling errors in individual lightcurves. From the other side,
the limb darkening effect in the transit curve is always symmetric relative to the
mid-time, implying that its impact on the derived mid-times should be small,
if not negligible. In fact, we noticed that even fitting
of a transit model without any limb darkening at all does not change the derived TTV
data significantly (beyond the estimated parametric uncertainties). However,
in the final version of our algorithm we decided to disentangle the limb darkening
coefficients for the best lightcurves (those that have r.m.s. smaller than $10$ per cent of
the transit depth) and fit these coefficients independently.

\item It is not taken into account that transit duration may also be subject to variations,
like the transit mid-time. However, in practice, it appeared that the accuracy
of the transit duration estimations was roughly an order of magnitude worse than those
of the mid-times. Moreover, some lightcurves cover the transit event only partially,
implying that its duration would remain almost unconstrained when fitted independently. We
do not address transit duration variations (TDV) in this work, focusing our attention on
the TTV.

\item The mid-time estimates, obtained in such a manner, are not necessarily uncorrelated.
These mid-times are correlated with the remaining transit parameters, which are now shared
between the lightcurves. Through this effect, some cross-correlation between the estimated
mid-times may appear. This creates a risk of unexpected statistical effects in the derived
TTV data, like e.g. the non-white noise. However, the magnitude of these TTV
correlations usually remains very small (maximum of a few per cent, and $\sim 0.1$ per cent
in average), and no deviations from the white noise are seen in the periodograms of the TTV
data. More significant ($>10$ per cent) TTV correlations can appear for lightcurves that
offer only a partial coverage of the transit, but such lightcurves represent
only a minor fraction of our data.
\end{enumerate}

In addition to the planetary transit itself, our lightcurve models include a
polynomial trend with fittable coefficients. Such a trend is necessary to take into account
various drifting effects, e.g. the effect of airmass or other types of
systematic variations that appear frequently in our data. Each transit lightcurve has an
individual fittable trend with a separate set of trend coefficients. After
some experimenting with the data, we found that cubic trends represent a good compromise
between the model adequacy and its parametric complexity. We did not try to reduce
this systematic photometric variation based on its correlation
with the airmass function: while for some lightcurves such a correlation looked clear, for
others it was not obvious, indicating that other systematic effects were in the game.

The lightcurves model is relatively complicated, while the photometry used in this work is
not of a very good quality. In practice we often faced difficulties causing the fitter
to be trapped in the local maxima of the likelihood function, which was related to
an unrealistic branch of the solutions. For example, without special care, see below, we
frequently obtained non-physical solutions corresponding to a grazing transit with $r\gg
1$. Also, some lightcurves do not cover the complete transit, and in such cases the fitting
of the transit mid-time is complicated by its strong correlation with the coefficients
of the polynomial trend. To avoid such traps, we worked out the following sequence
of auxiliary preliminary fits:
\begin{enumerate}
\item Perform a preliminary fit constraining the mid-times at a regular grid (with
free scale and offset), implying all TTV residuals are zero by definition;
fixing the impact parameter $p$ at an intermediary value of $1$ ($b\approx 1/\sqrt 2$), and
fixing the limb darkening coefficients at $A=B=0.25$ (or
$\theta=\pi/4$ and $\varphi=\pi/3$).
\item Refit after releasing the mid-times and impact parameter, but still holding the
limb darkening coefficients fixed.
\item Refit after releasing the limb darkening coefficients (binding them across different lightcurves).
\item Refit after full release of the limb darkening coefficients for the best lightcurves
(those with r.m.s. $<0.1$ of the transit depth).
\end{enumerate}

After that, we also apply the red-noise detection and fitting procedure as described in the
section below.

\subsection{Reduction of the red noise}
\label{subsec_an_rn}
It is already known well that stellar photometric data usually include a
correlated (``red'') noise component that has an important effect on the
fitted transit parameters \citep{Pont06,Winn08,CarterWinn09}. An easy technique to
calculate the effect of the red noise on the fitting uncertainties was introduced in these
works. Although these authors avoid making restrictive assumptions
concerning the correlation structure of the red noise, their method still remains rather
simplistic and it does not take into account important effects. They mainly focus on a more
accurate determination of the uncertainties in the transit
fit, which are typically underestimated without a proper treatment of the red noise. But
the red noise may also induce biases in the fitted parameters themselves. This biasing
effect was already noted when processing Doppler data affected by red noise in a generally
simialar way \citep[e.g.][]{Baluev11,Baluev13a}. Also, there is not an obvious way
to control the validity of the underlying assumptions in the traditional approach of the
photometric red noise reduction. This approach does not offer a complete noise model
that could be verified for accuracy. Additionally, the original method by
\citet{Pont06} relies on the out-of-transit photometric data, which are very limited in our
case.

In this work we apply the approach based on a parametric modelling of the noise covariance
matrix. This is an adaptation of the red-noise fitting technique from
\citep{Baluev11,Baluev13a,Baluev14a} that was developed to handle exoplanetary
radial-velocity data, in which the noise correlations may appear e.g. due to the stellar
activity. In this approach we approximate the red noise by a stationary Gaussian
random process with a covariance function of a given functional form. The covariance
coefficient between two arbitrary photometric observations $x_i$ and $x_j$, acquired at the
times $t_i$ and $t_j$, is modelled as
\begin{eqnarray}
V_{ij} = \cov(x_i,x_j) = \sigma_{i,\rm wht}^2(p_{\rm wht})\delta_{ij} + V_{ij,\rm red}(p_{\rm red},\tau), \nonumber\\
V_{ij,\rm red} = p_{\rm red} R_{ij}(\tau), \quad R_{ij} = \rho\left(\frac{t_i-t_j}{\tau}\right),
\label{rednoise}
\end{eqnarray}
with $p_{\rm wht}$, $p_{\rm red}$, and $\tau$ being free
fittable parameters. Here, the first noise term, $\sigma_{i,\rm wht}^2$,
represents the white fraction of the noise, detailed below. The red noise is given
by the second term, $V_{ij}$, which also depends on the function $\rho(t)$ that represents
an adopted shape of the correlation function. We use
mainly the exponential correlation function $\rho(t)=\exp(-|t|)$. This not a unique choice,
but in practice this model of the red noise usually appears adequate.

The white-noise term $\sigma_{i,\rm wht}^2$ in~(\ref{rednoise}) requires a
separate discussion. We believe that physically the so-called additive model would be more
suitable here. In this model, the value $\sigma_{i,\rm wht}^2$ is determined as a sum of
the stated instrumental variance and of an unknown fittable ``jitter'' variance, generated
by e.g. Earth atmospheric instability, or by unassessed instrumental instability, or
by short-term (minutes to hours) intrinsic variations of the stellar flux,
whenever these variations can be treated as a white noise. Although, we must note that
in practice the instrumental uncertainties in our data do not look very trustable, and
often they are just omitted, forcing us to assume that the measurements have
just equal uncertainties. Therefore, in any case we should not expect that our white noise
can be accurately modelled from the physical point of view. In these
circumstances, the noise model should be chosen mainly on the basis of
mathematical simplicity or usefulness, relying on only minimal or no knowledge
of the underlying physics. We decided to use in our work a regularized noise model defined
in \citep{Baluev14a}. In the majority of practical cases, this model should be
equivalent to the additive noise model. This regularized model proved rather resistant with
respect to various pitfalls appearing during the fitting procedure.

The fitting of the compound model, involving simultaneously the models of the transit curve
and of the photometric noise, is done by means of the maximum-likelihood approach with
details given in \citep{Baluev14a}. That work was devoted primarily to the analysis of the
radial velocity data, but mathematically the methods that we are using here are identical.

However, after an attempt to apply this technique in practice literally,
we faced the problem that the red noise could not be fitted in many of our
lightcurves. Only $1/3$ to $1/2$ of the lightcurves allowed for a reliable red noise
fit, while in the other cases, the red noise was either not detectable, or its estimated
magnitude became pretty small (in comparison with the uncertainty). Such cases lead to
model degeneracies and decrease the reliability of the
entire fit for a given star. Thus, we decided to add the red noise term to the photometric
model only in the cases when it was justified.

We added to our analysis pipeline a set of auxiliary fits in order to identify the
lightcurves, in which the red noise could be modelled more or less reliably. Namely,
we tried to perform a series of test fits, latterly adding a red noise term to the model of
each transit lightcurve. If the red noise could not be estimated reliably, more accurately
if the relative uncertainty of its estimated magnitude exceeded $2/3$, the
relevant red noise term was removed from the model and this lightcurve was further treated
as having purely white noise. Otherwise, we proceeded to the next test fit preserving this
red noise term in the model. In the end, after all individual lightcurves were processed in
such a way, we performed one more check of each red noise term (because some of them could
turn insignificant after the other terms being added) and the final fit was made.

\begin{figure*}
\includegraphics[width=0.99\textwidth]{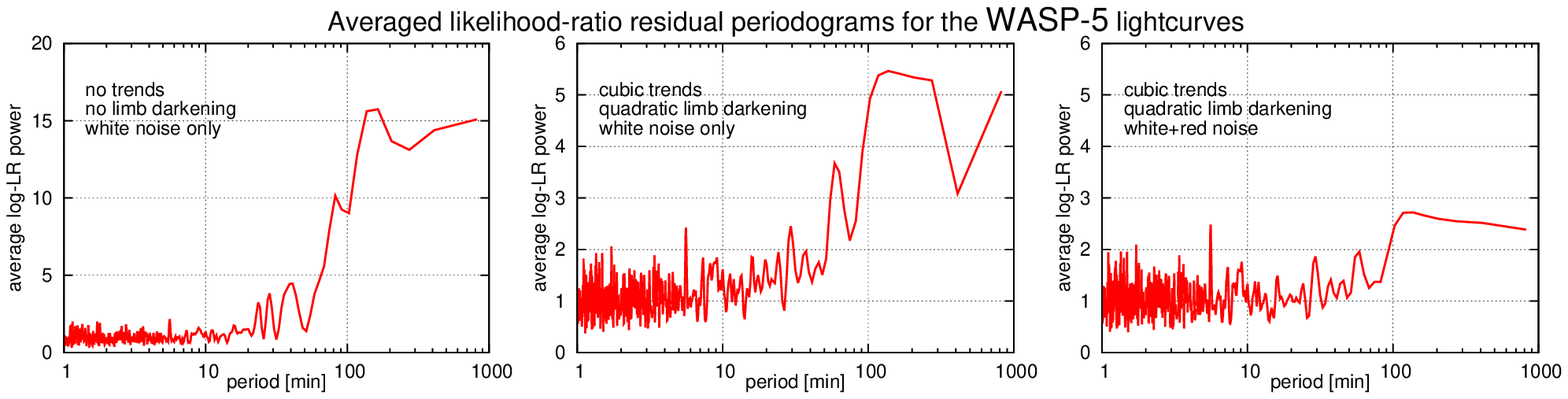}\\
\includegraphics[width=0.99\textwidth]{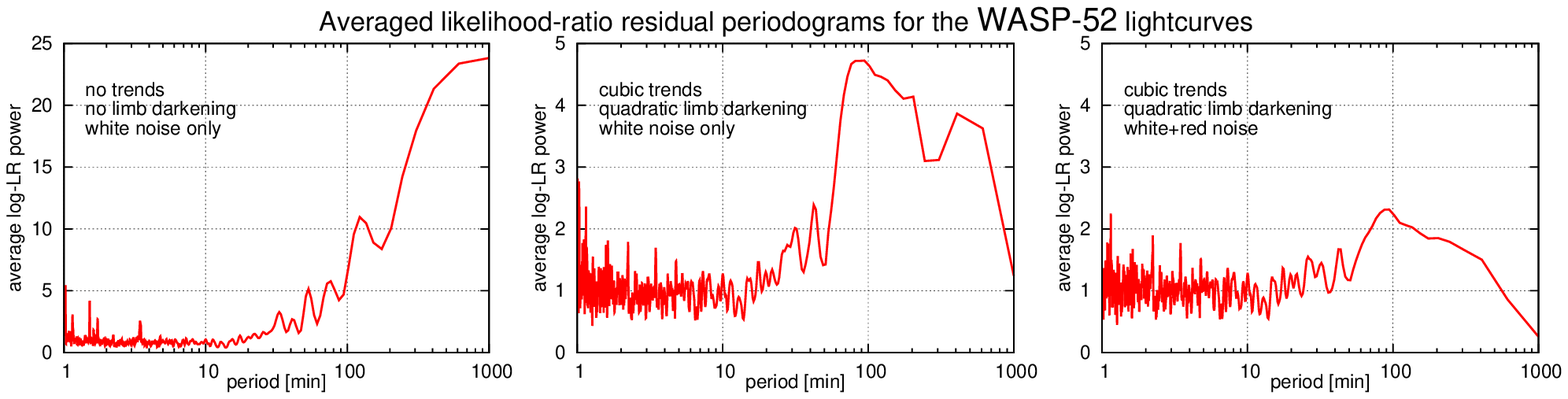}\\
\includegraphics[width=0.99\textwidth]{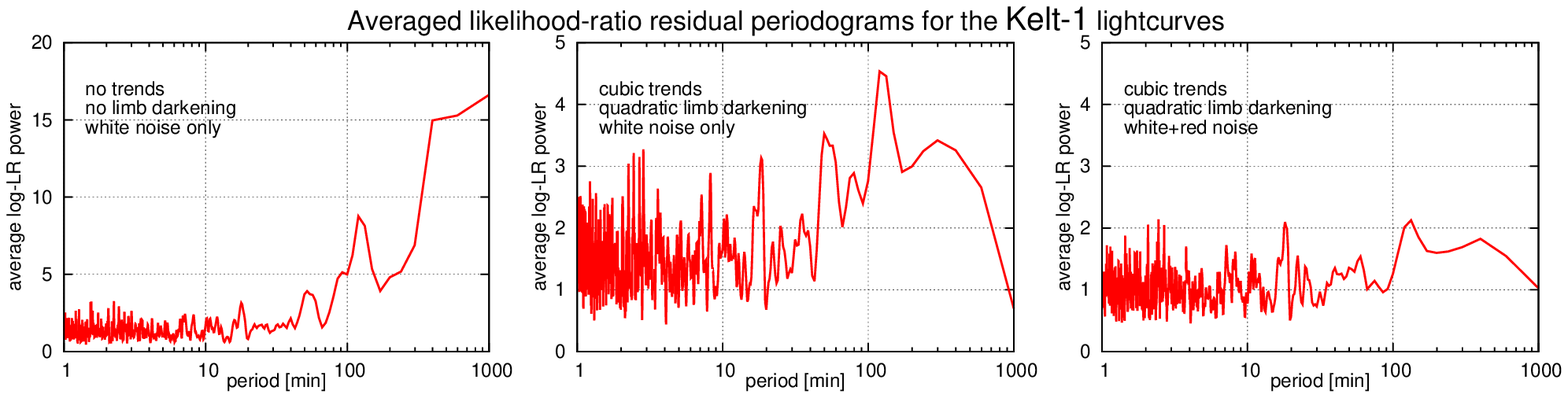}\\
\includegraphics[width=0.99\textwidth]{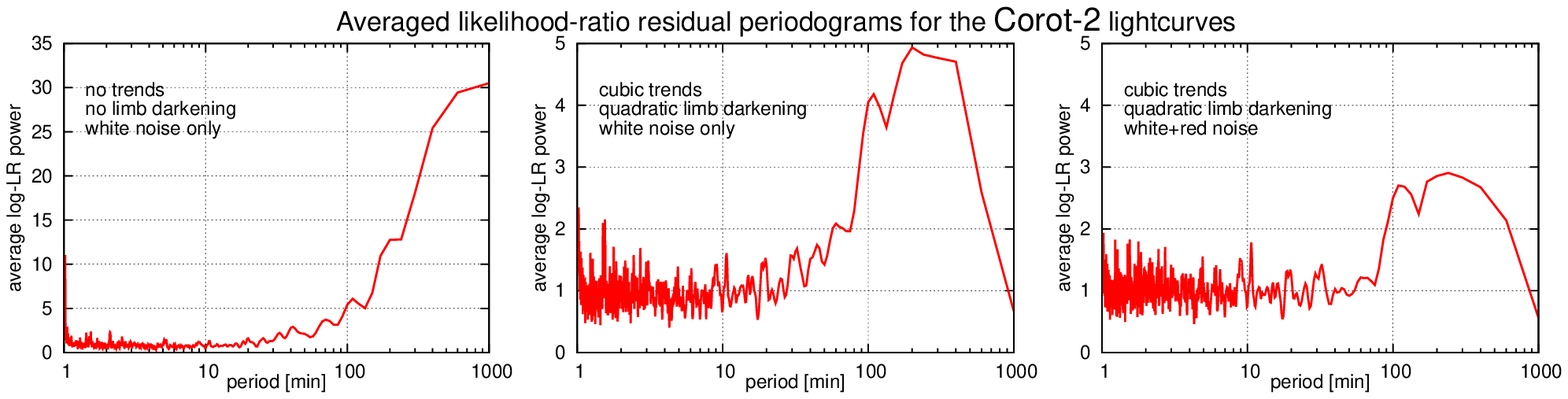}\\
\includegraphics[width=0.99\textwidth]{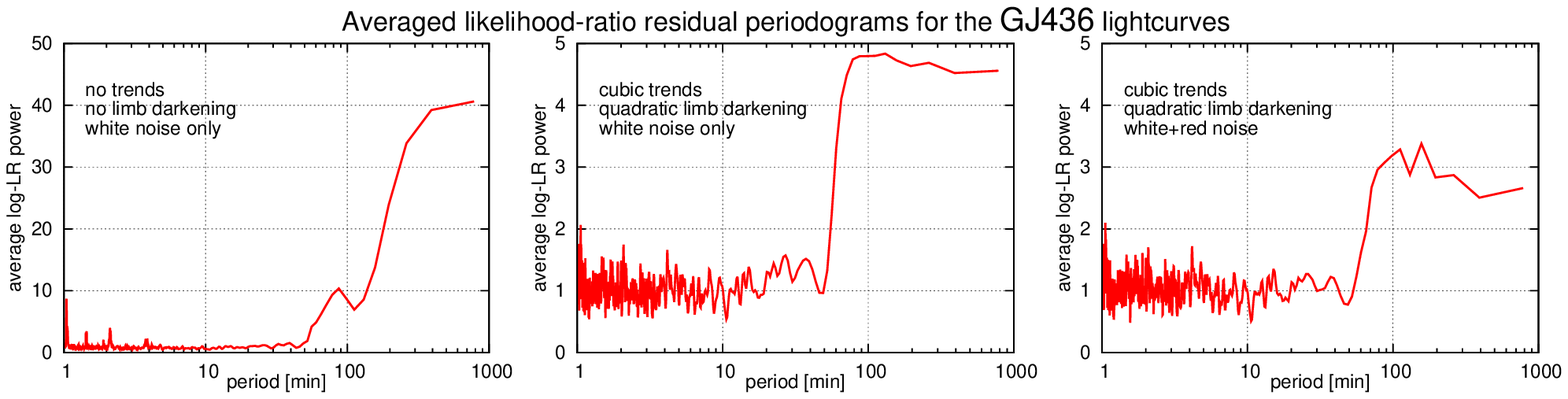}
\caption{Examining the effect of the red noise in the lightcurve photometry for a few of stars
belonging to our sample. Please note that plots in the first column have different ordinate
scale than those in the other two. See Sect.~\ref{subsec_an_rn} for a detailed discussion.}
\label{fig_rsdpow}
\end{figure*}

The quality of such an approach to the red noise reduction is examined
in Fig.~\ref{fig_rsdpow}. The best way to observe a ``non-white'' noise is to look
at its power spectrum: the white noise should have constant power in average, while the red
noise should demonstrate a systematic uprise to long periods. In our case we compute the
so-called residual periodograms \citep[see][]{Baluev14a} for each individual lightcurve,
and then compute their average per each star involved in the analysis. These
averaged periodograms are plotted in Fig.~\ref{fig_rsdpow}. For each star, we sequentially
try three base photometric models: (i) just transits without any limb darkening plus the
white noise; (ii) same as model (i) plus cubic trends and the effect of the quadratic limb
darkening; (iii) same as model (ii) plus the red noise term.

We can clearly see from Fig.~\ref{fig_rsdpow} that the long term photometric variations are
not reduced to just trends. Although the trends themselves are also important, after their
removal the power spectra still contain an additional power in the period range of
$>10$~min. The trends can only affect the period domain of $>200-300$ min (this corresponds
to the typical time span of the lightcurves). In the most cases, application
of our red-noise model remarkably suppresses the remaining excessive power, at least
by a factor of $2$. We must acknowledge that the reduction of this red noise looks
rather difficult and still not entirely complete.

Some statistical information on the derived red noise characteristics now follows. About
$60$ per cent of lightcurves did not contain any detectable red noise at
all. For the remaining $40$ per cent of lightcurves, the red noise caused an increase of
the TTV uncertainty by a factor of $1.3$ on average (a median value). This is not very
large, although a minor fraction of the mid-times demonstrated an increase
in the uncertainty up to a factor of $2-3$. Concerning the bias, appearing
in the derived mid-times themselves (rather than uncertainties), its average value was only
about $1/6$ fraction of the uncertainty, and only a few points
demonstrated a shift above $1\sigma$.

Summarizing, in the majority of the cases, either there was not any detectable red
noise, or the change in the TTV data was small, even if the red noise term was detectable.
Nonetheless, it is important that some individual TTV points are affected much more.

\subsection{Testing the Gaussianity of the noise and clearing away the outliers}
\label{subsec_an_gauss}
\begin{figure*}
\includegraphics[width=\linewidth]{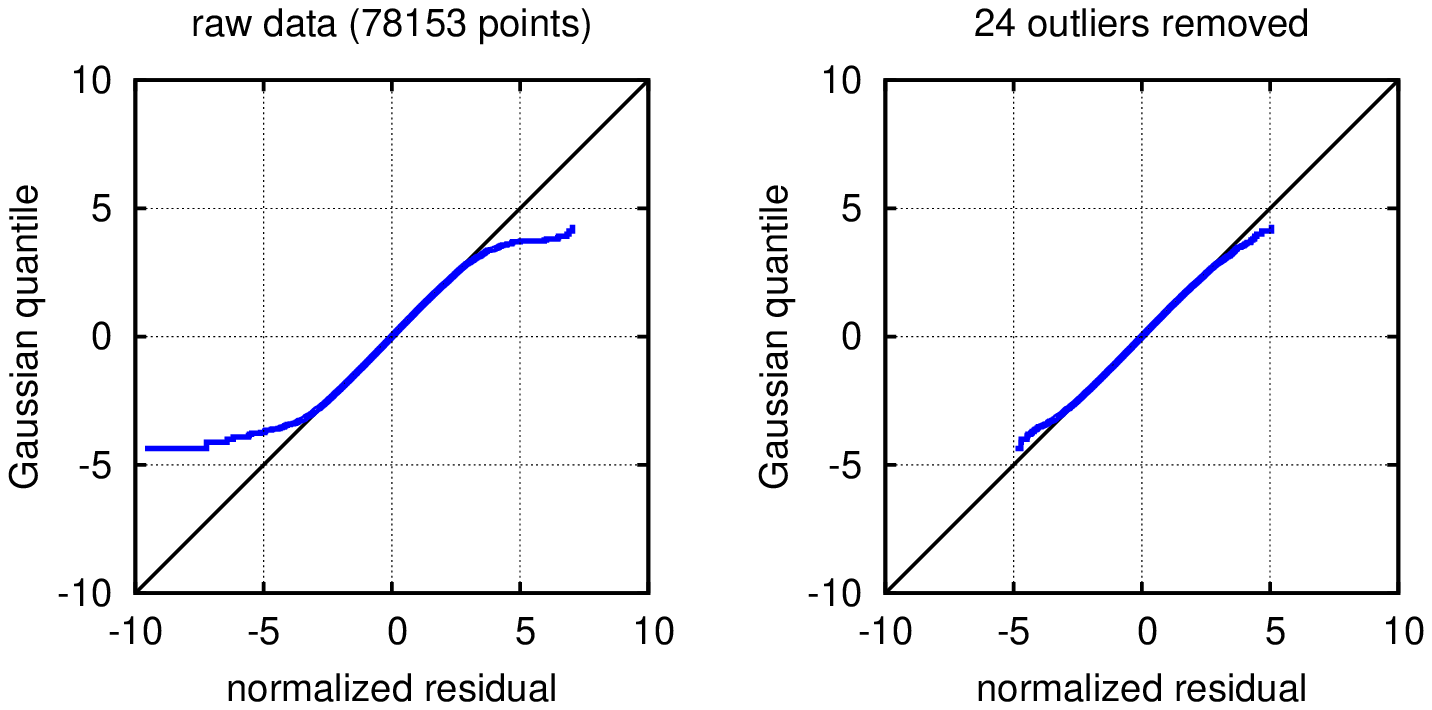}
\caption{Testing the normality of the entire $\sim 80000$ photometric data used in the
work. Abscissa shows the value of the best fitting residual, normalized by its total
modelled variance (for white+red noise). The ordinate contains the normal quantiles of the
empirical cumulative distribution of these normalized residuals. Perfectly
Gaussian residuals should all lie close to the main diagonal that represents a
standard normal distribution, while a deviation from the diagonal indicates a non-Gaussian
noise.}
\label{fig_gauss}
\end{figure*}

In the data-analysis methods described above, we largely relied on the assumption that the
photometric noise is Gaussian. This assumption needs some verification, which we performed
in the following manner. After fitting all the data, we computed the residuals and
normalized them by their relevant modelled uncertainty (square root of the sum
of the estimated white and red noise variances).
If the noise was Gaussian, the distribution of these normalized residuals should
be close to the standard Gaussian, and vice versa. We consider the entire set of $\sim
80000$ normalized residuals for all stars involved in our analysis, and plot the
relevant distribution in Fig.~\ref{fig_gauss}

As we can see, the empirical distribution is indeed almost Gaussian, except for the very
tails ($>4$-sigma deviation). The empirical distribution has heavier tails. However,
by removing data in the distribution tails we can reach a good agreement with the Gaussian
approximation. In fact, these extreme points represent outliers and should be removed in
any case. The final results below correspond to the data with these $24$ outliers removed.
We also notice that the minor systematic deviations still remaining in the
distribution tails are mainly due to \citet{Bakos06} data for HD~189733.
Visual investigation reveals that some of these lightcurves contain limited segments
with unexpectedly increased photometric scatter. Removal of the \citet{Bakos06} data makes
the agreement of the residuals distribution with the Gaussian one very good.

\section{Derived TTV data and their analysis}
\label{sec_ttv}
Using the fitting techniqe described above, we computed the mid-times for each
observed transit, and placed these TTV measurements and other accompanying data in the
online supplement to the article. The supplement represents a file containing a text table.
The meaning of the columns in this file is described in Table~\ref{tab_TTV}. This file only
contains the parameters that are attached to individual lightcurves. The
common fit parameters that are shared between different lightcurves are given separately in
Sect.~\ref{sec_trpar} below.

\begin{table}
\caption{Explanation of the TTV data file}
\label{tab_TTV}
\begin{tabular}{cp{7cm}}
\hline
column & explanation   \\
\hline
\multicolumn{2}{c}{Primary data}\\
1      & integer transit count (number of the transiter's revolutions, restarts from zero for each star) \\
2, 3   & fitted transit mid-time (${\rm BJD}_{\rm TDB}$) and its uncertainty \\
\hline
\multicolumn{2}{c}{Auxiliary data (other parameters of the fit)}\\
4      & a standardized name of the lightcurve file that includes the date, the target name, and the name of the observer or first author of a paper \\
5      & the reference time $T_0$ to which the following trend coefficients refer \\
6      & number of the following trend coefficients including the constant, i.e. the trend degree plus one (always $4$ in this work) \\
7, 8   & fitted constant level of the magnitude and its uncertainty \\
9, 10  & fitted linear trend coefficient and its uncertainty (mag$\cdot$day$^{-1}$)\\
11, 12 & fitted quadratic trend coefficient and its uncertainty (mag$\cdot$day$^{-2}$)\\
13, 14 & fitted cubic trend coefficient and its uncertainty (mag$\cdot$day$^{-3}$)\\
15, 16 & fitted limb darkening coefficient $A$ and its uncertainty\\
17, 18 & fitted limb darkening coefficient $B$ and its uncertainty\\
19     & an adopted value of a scale parameter $\sigma_{\rm scale}$ needed to fully characterize the regularized model of the photometric noise, see \citep{Baluev14a} \\
20, 21 & fitted photometric white jitter $\sigma_{\star,\rm wht}$ (def. in \citet{Baluev14a}) and its uncertainty\\
22, 23 & fitted photometric red jitter $\sigma_{\star,\rm red}=\sqrt{p_{\rm red}}$ and its uncertainty\\
24, 25 & fitted correlation timescale $\tau$ of the red jitter (days)\\
26     & r.m.s. of the best-fit residuals for this lightcurve\\
\hline
\end{tabular}
\end{table}

\begin{figure*}
\includegraphics[width=0.49\textwidth]{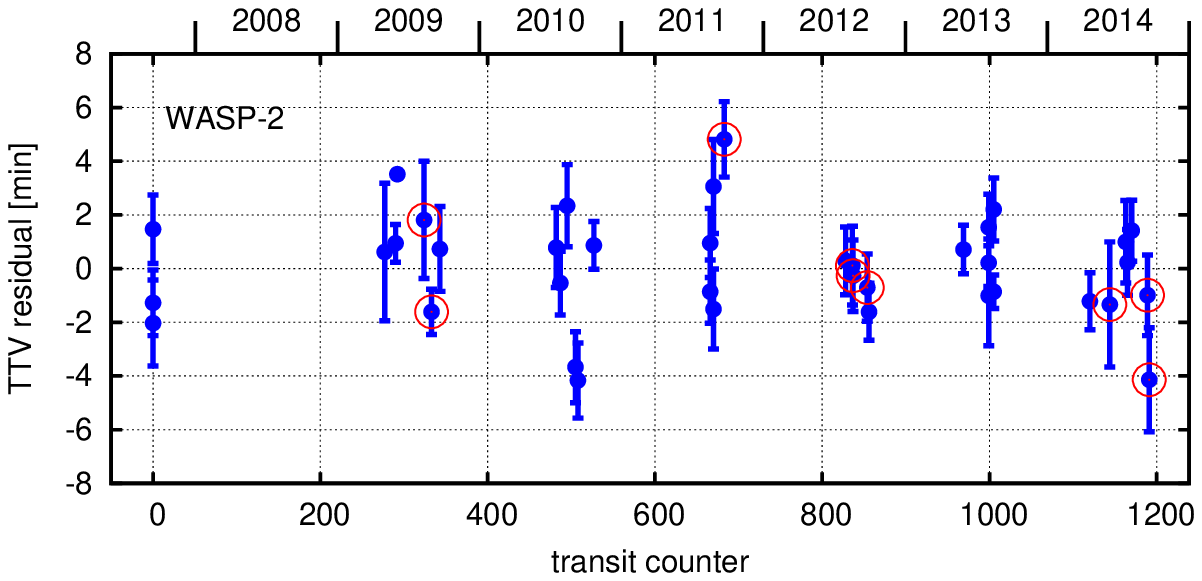}
\includegraphics[width=0.49\textwidth]{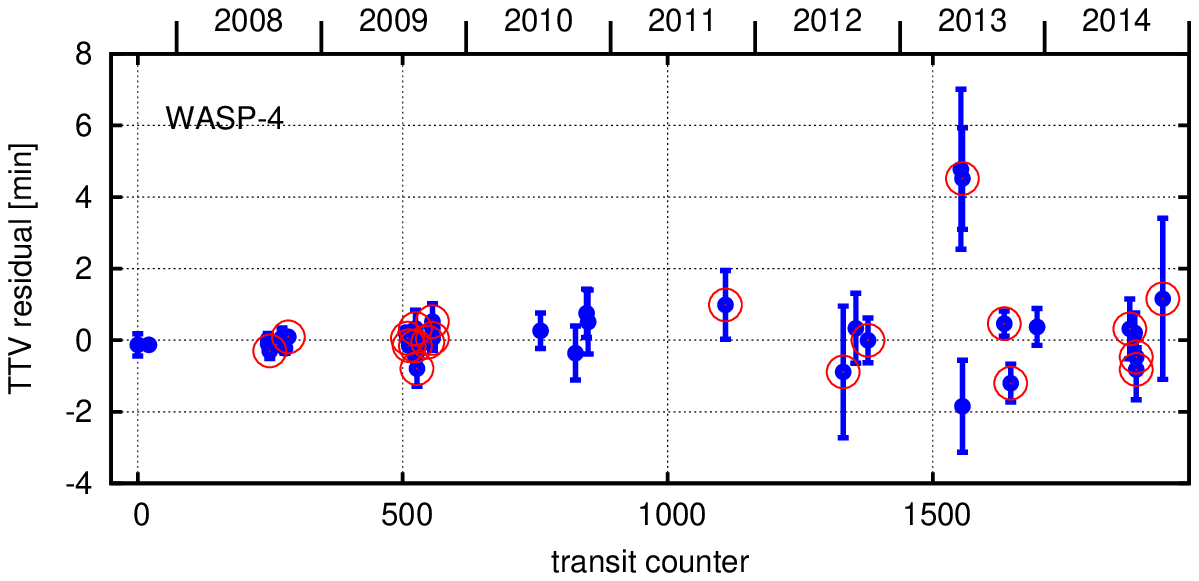}
\includegraphics[width=0.49\textwidth]{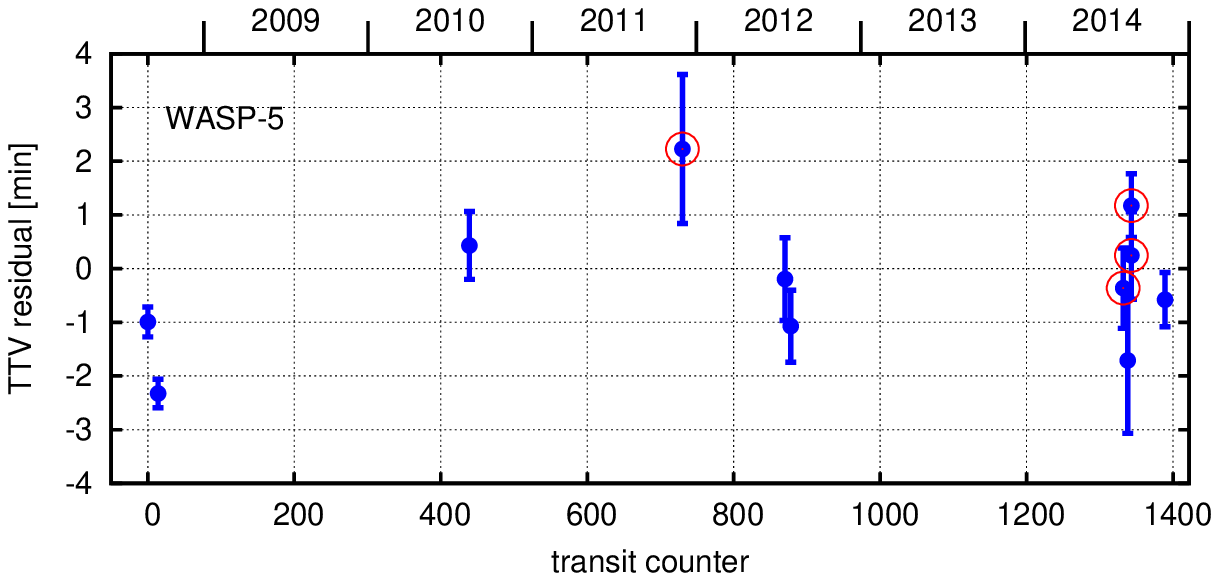}
\includegraphics[width=0.49\textwidth]{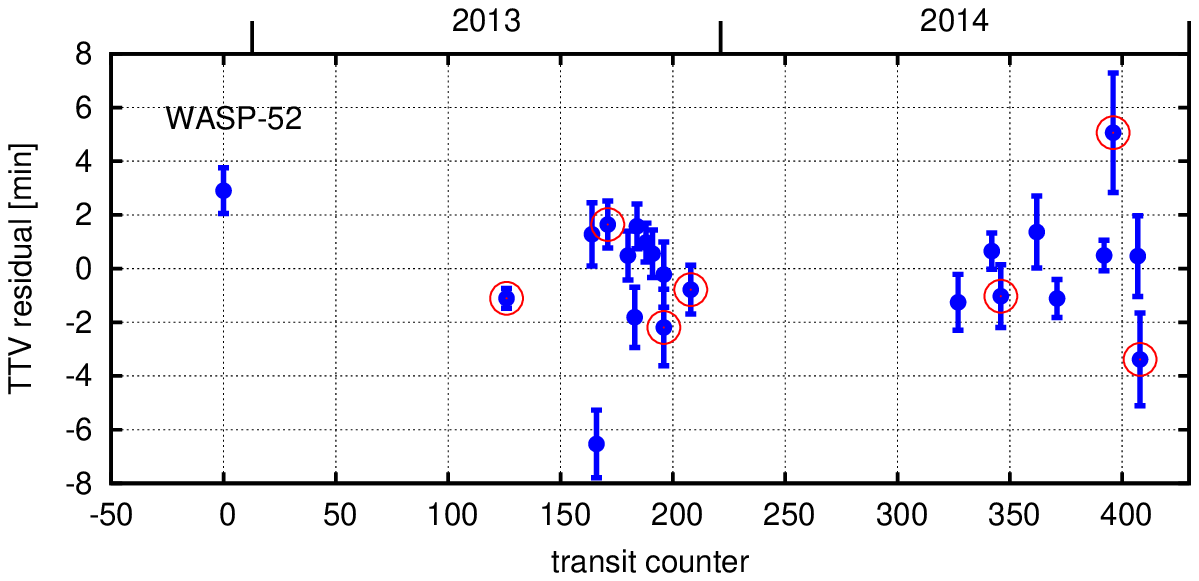}
\includegraphics[width=0.49\textwidth]{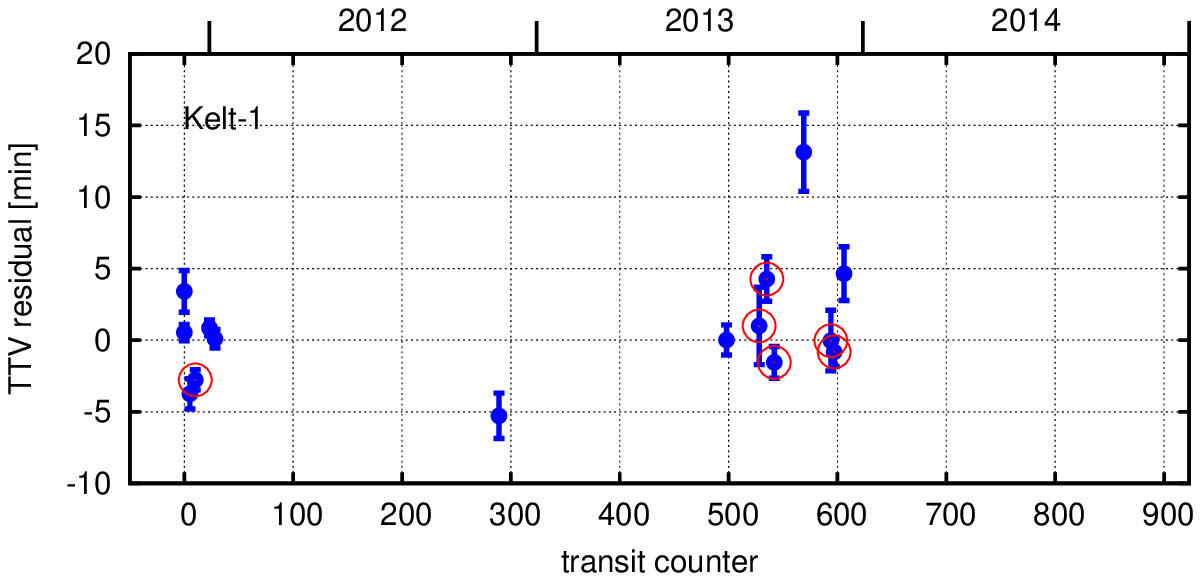}
\includegraphics[width=0.49\textwidth]{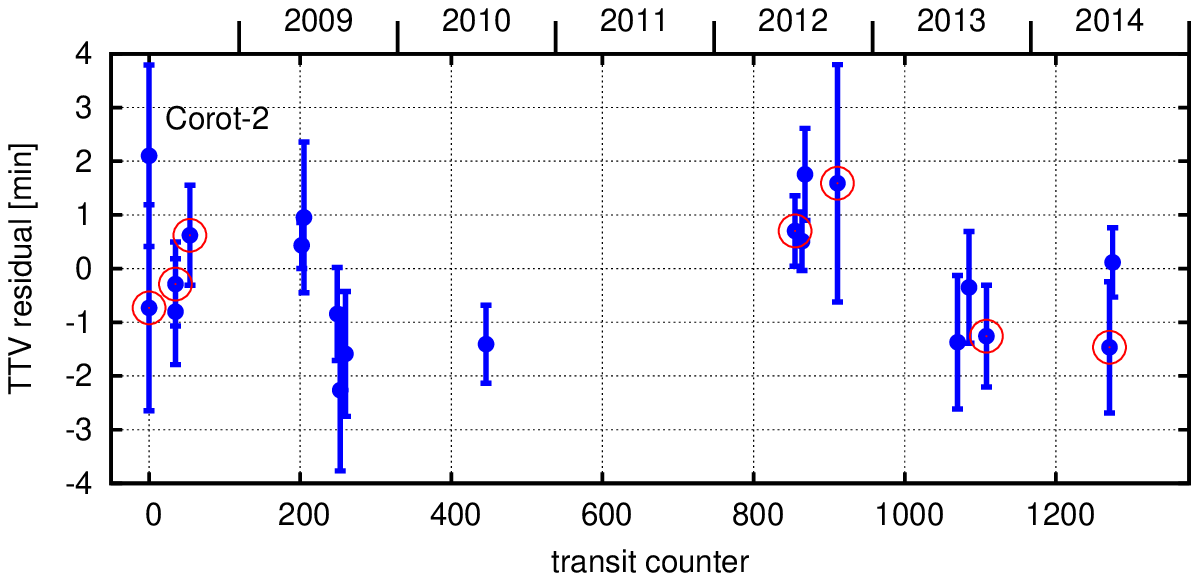}
\includegraphics[width=0.49\textwidth]{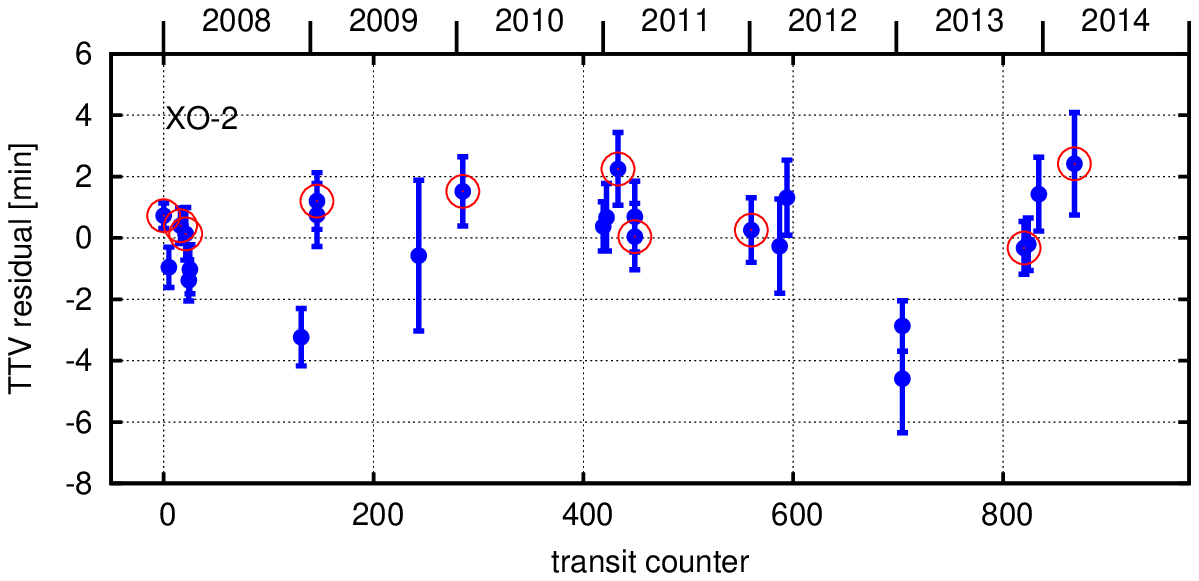}
\includegraphics[width=0.49\textwidth]{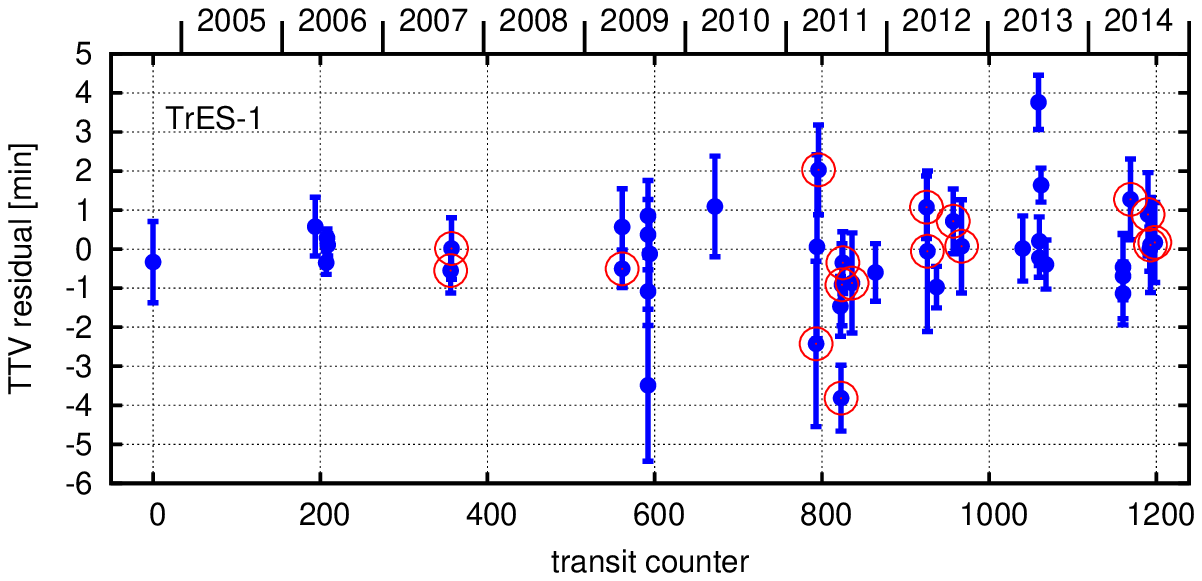}
\includegraphics[width=0.49\textwidth]{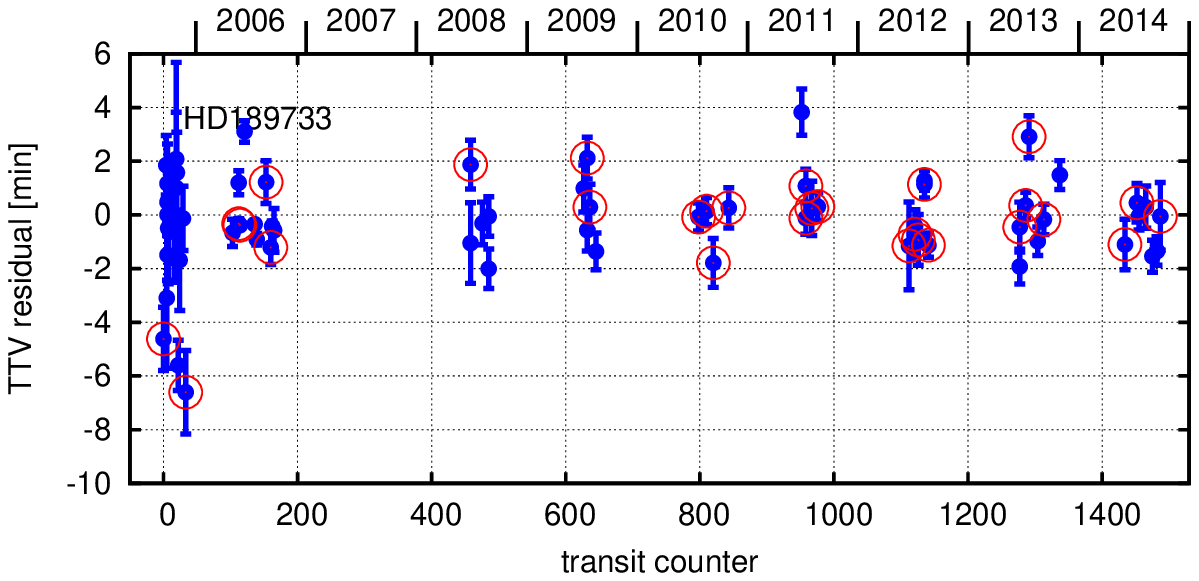}
\includegraphics[width=0.49\textwidth]{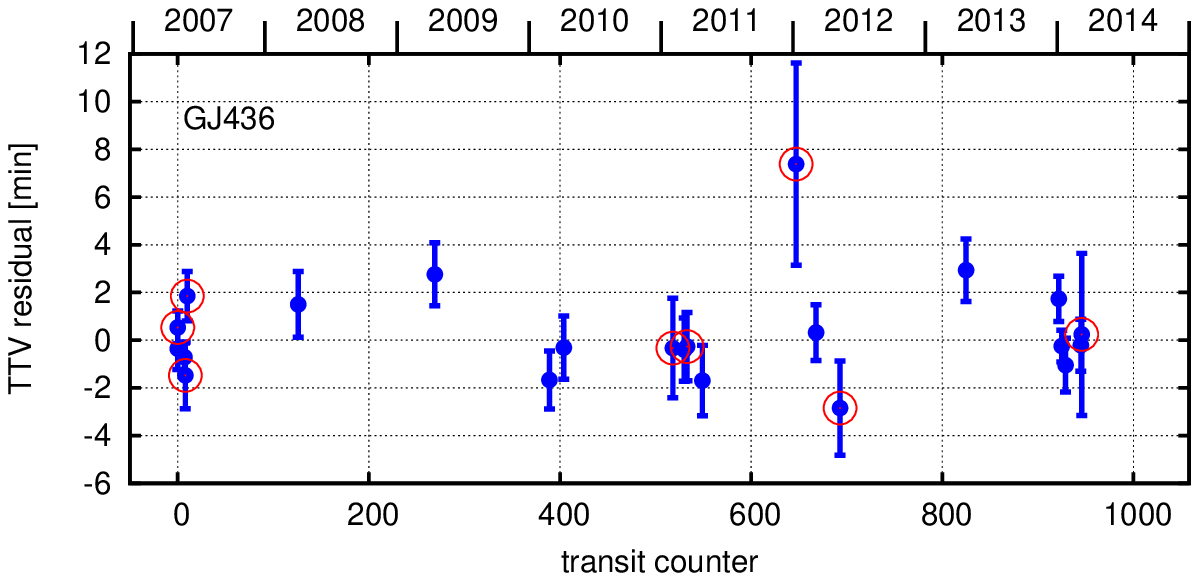}
\caption{Graphs of the TTV residuals, based on the
best fitting transit ephemeris (values of $P$ and $t_c^{\rm ref}$)
from Table~\ref{tab_traw}. The transits for which a significant red noise was detected
in the photometry are labelled with a circle.}
\label{fig_ttv}
\end{figure*}

\begin{figure*}
\includegraphics[width=0.99\linewidth]{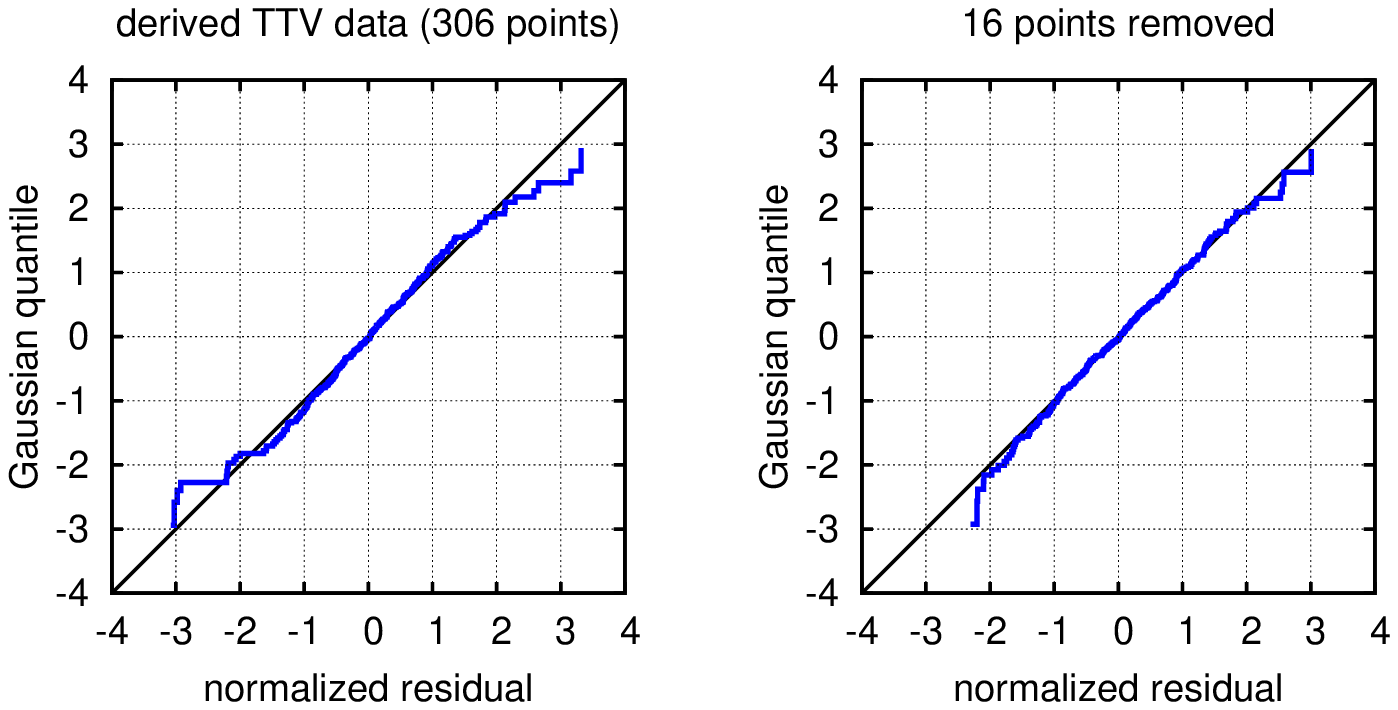}
\caption{Testing the normality of the TTV data derived in the
work. The plot is similar to Fig.~\ref{fig_gauss}, but refers to the TTV residuals
(relatively to a strictly periodic transit ephemeris) rather than to the photometric ones.
In the right panel, we removed the \citet{Southworth09a,Southworth09b,Southworth10} data
($7$ points), as well as a few potential outliers ($9$ points).}
\label{fig_ttvgauss}
\end{figure*}

The TTV residuals for these data are plotted in Fig.~\ref{fig_ttv}. In
Fig.~\ref{fig_ttvgauss} we verify the normality of the distribution of the TTV noise
present in these derived data in the way similar to Sect.~\ref{subsec_an_gauss}. In the TTV
residuals we find no obvious outliers or deviations from the
Gaussian distribution. The apparent excess in the distribution tails that can be seen in
the left panel of Fig.~\ref{fig_ttvgauss} is uncertain due to a relatively small number of
TTV data points. Nonetheless, we identified $9$ possible ``candidate outliers'', and tried to
remove them together with $7$ TTV points referring
to the \citet{Southworth09a,Southworth09b,Southworth10} data. The TTV analysis discussed
below was performed for the full TTV data set as well as for the reduced one.

First, we made an effort to detect possible long-term curvature in our TTV. This was
done by means of modelling the mid-times by a quadratic function (instead of only a linear
function that would correspond to strictly periodic transits). None of the TTV time series
demonstrated a long-term quadratic variation with at least $2$-sigma significance (the most
suspicious cases had $1.5\sigma$ to $1.7\sigma$).

Second, we undertook a periodorgam analysis of the TTV residuals, in order to reveal
possible periodic perturbations from any additional unknown planets orbiting these
stars. We again applied ``residual periodograms'' similar to the ones that were used
to plot Fig.~\ref{fig_rsdpow} above. Now our null model included a quadratic trend that is
related to the period of the main (transiting) planet and white noise with a
fittable additive ``jitter''. The alternative model also included a sinusoidal signal at a
probe frequency. None of these periodograms demonstrated any signs of a non-white noise,
and in fact almost all appeared consistent with the pure noise model, although the cases
revealing significant excessive TTV ``jitter'' were rather frequent.
No significant TTV periodicities were found in the data for any of the stars involved in
the analysis, except for the WASP-4 case which is discussed below. The periodogram
significance levels were estimated using the approach of \citet{Baluev08a,Baluev08b}.

We believe that this result indicates a considerable improvement in the statistical quality
of the TTV data and/or methods of the statistical analysis, because for the original TTV
data from ETD we frequently detected spurious periodicities possibly appearing due
to imperfections of the transit fitting algorithm.

\section{Improved transit curve parameters}
\label{sec_trpar}
In addition to the TTV data, we also provide the remaining best fitting parameters
of the exoplanetary transit lightcurves. These are the transit parameters that were shared
between different lightcurves, and they are given in Table~\ref{tab_traw}. In addition
to the previously mentioned transit parameters $r$, $t_d$, $b$, and the limb darkening
parameters $A$ and $B$, this table also contains the following data: number of the transit
lightcurves used in the analysis, number of the lightcurves in which a red noise
term was robustly detected, the $\chi^2$ (weighted r.m.s.) of the TTV residuals (relatively
to the best linear fit of the derived mid-times), and the refined parameters of the transit
emphemeris: transiter period $P$ and a reference mid-time $t_c^{\rm ref}$. The
last two quantities were obtained by requiring the mid-times to obey a strict linear
relation with the transit count and fitting the coefficients of this relation. The value of
$t_c^{\rm ref}$ is not unique due to the periodic nature of the transits. We chose
such a reference transit, for which the uncertainty of $t_c$ would be minimum.
This condition simultaneously implies that the correlation between the estimated $P$ and
$t_c^{\rm ref}$ is zero. Note that it is not required to have actual transit observation at
this $t_c^{\rm ref}$.

The $\chi^2_{\rm TTV}$ values in Table~\ref{tab_traw} always exceed unity,
so the fit uncertainties might be moderately underestimated. In some cases this higher than
expected scatter could be partly explained by small number of transits involved in the
analysis, and in some part it might be due to additional unseen planets that may
induce complicated TTV signals and are difficult to detect. Another explanation is
incomplete reduction of the red noise and the effect of nonlinearity of the transit model.
The latter effect can introduce biases both in the estimated parameters as well as in their
uncertainty estimations. So far we could not decide which explanation is
more likely. Interestingly this excessive TTV scatter persists and is important even
in such a robustly fittable case like HD~189733, in which the model nonlinearity should be
well suppressed thanks to a large number of available observations and lightcurves.

\begin{landscape}
\begin{table}
\caption{Fitted parameters of exoplanetary transit curves}
\label{tab_traw}
\begin{tabular}{@{}l@{}cclllccc@{}llp{3.5cm}@{}}
\hline
          & total      & \multicolumn{6}{c}{Assuming fittable TTV}& &\multicolumn{2}{c}{Fixing TTV residuals at zero$^1$} &  \\
\cline{3-8}\cline{10-11}
transiter & number of  & number of & radii ratio & half-duration& impact par.& & mid-times correl. & & orbital period & ref. mid-time  & comment \\
          & transits   & red-noised & $r=R_{\rm pl}/R_\star$ &$t_d/2$ [days]& $b$         & $\sqrt{\chi^2_{\rm TTV}}$ $^1$& MAD/MAX & & $P$ [days]&$t_c^{\rm ref}$ [${\rm BJD}_{\rm TDB}-$&              \\
          &            & transits &  & & & & & & &$-2450000$]&              \\
\hline
WASP-2~b  & $38$       & $9$       & $0.1355(31)$ & $0.03701(35)$ & $0.7380(94)$& $1.38$ & $0.00056/0.0026$ & & $2.15222163(42)$ & $5894.07919(15)$ &  \\
WASP-4~b  & $43$       & $20$      & $0.15495(32)$& $0.044898(53)$& $0.136(28)$ & $1.35$ & $0.00025/0.0039$ & & $1.338231624(68)$& $4966.782814(21)$ & \\
WASP-5~b  & $11$       & $4$       & $0.1136(13)$ & $0.05028(29)$ & $0.446(40)$ & $1.34$ & $0.0070/0.10$ & & $1.62842953(52)$ & $6446.98868(17)$ & \\
WASP-52~b & $22$       & $7$       & $0.1629(44)$ & $0.03858(69)$ & $0.598(32)$ & $2.00$ & $0.0020/0.016$ & & $1.7497835(11)$  & $6673.82149(13)$ & \\
Kelt-1~b  & $15$       & $6$       & $0.0783(14)$ & $0.05661(42)$ &$0.998(44)^2$& $2.56$ & $0.00092/0.017$ & & $1.21749448(80)$ & $6093.13464(19)$ & Showing $\sqrt{1-b^2}$ instead of $b$, see note~2 \\
Corot-2~b & $20$       & $7$       & $0.1639(21)$ & $0.04736(36)$ &$1.000(89)^2$& $1.17$ & $0.00079/0.015$ & & $1.74299673(31)$ & $5628.44758(14)$ & Showing $\sqrt{1-b^2}$ instead of $b$, See note~2 \\
XO-2~b    & $25$       & $10$      & $0.1036(13)$ & $0.05563(32)$ &$0.996(31)^2$& $1.56$ & $0.00056/0.012$ & & $2.61585779(43)$ & $5139.16092(13)$ & Showing $\sqrt{1-b^2}$ instead of $b$, See note~2 \\
TrES-1~b  & $43$       & $17$      & $0.13781(97)$& $0.05225(18)$ & $0.191(67)$ & $1.54$ & $0.00038/0.0078$ & & $3.03006973(18)$ & $5016.969937(70)$ & \\
HD189733~b& $67$       & $29$      & $0.15712(40)$& $0.037576(48)$& $0.6636(19)$& $2.16$ & $0.00037/0.071$ & & $2.218575200(77)$ & $3955.5255511(88)$& Some of \citet{Bakos06} data contain many outliers and possibly non-Gaussian noise \\
GJ436~b   & $22$       & $8$       & $0.088(10)$  & $0.02102(71)$ & $0.806(31)$ & $1.22$ & $0.00026/0.0019$& & $2.64389846(44)$ & $5280.17568(17)$ & Most data are of poor quality (large trends), and a significant eccentricity of $\sim 0.15$ is not taken into account\\
\hline
\end{tabular}\\
$^1$The orbital periods, the reference mid-times, and the values of $\chi^2_{\rm TTV}$ were
obtained after removal of the \citet{Southworth09a,Southworth09b,Southworth10} data,
as these data are affected by clock errors.\\
$^2$Impact parameter $b$ is close to zero and is thus a highly nonlinear parameter here.
Its estimation is severely nongaussian, and the formal uncertainty $\sigma_b$ is
much larger than $b$. To handle this pecularity, the values in the
column for $b$ are replaced by $a=\sqrt{1-b^2}$ and its uncertainty $\sigma_a = (b/a) \sigma_b$,
which are more informative here.
\end{table}
\end{landscape}

\section{RV data and their analysis}
\label{sec_RV}
In addition to the TTV time series from the photometry, we also derive RV data for some of
our stars using spectra found in the HARPS, HARPS-N, and SOPHIE archives. We process these
spectra using the advanced HARPS--TERRA pipeline by \citet{AngladaEscudeButler12}.
Public spectra were available for the following targets: WASP-2, 4, 5,
HD~189733, GJ~436, CoRoT-2.

Most of these RV data appeared less accurate
than the typical $1$~m/s precision demonstrated by HARPS. This is because many
of these targets are rather faint. Moreover, it seems that adequate modelling of these data
should involve non-trivial treatment of the RV noise. Frequently, these data are
combined in short series acquired within a few hours or even shorter. The data within such
single-night series should be significantly correlated \citep[e.g.][]{Nelson14}, and
this effect can be easily detected in some of our data. Usually, these short-term runs were
clearly intended to catch the Rossiter-McLaughlin effect during the planetary transit.
On larger time scales (days to weeks), the RV noise may still remain correlated as well,
likely due to the stellar activity \citep[see e.g.][]{Baluev13a,Robertson14}, and this type
of correlation is different from the one emerging at short time scales.

The full modelling of all these effects is outside of the scope of this paper, which was
intended to deal mainly with the photometry and TTV data. However, these RV data may carry
useful information that can be helpful in verifying the results of our TTV analysis.
In this work we adopted a simplified ``first look'' approach to the RV
data analysis. First, we replaced the RV series acquired in a single night
by their averages. This eliminated the need to model the Rossiter-McLaughlin effect,
as well as the RV noise correlations appearing within a single night. Of course,
such a procedure likely adds some systematic error to these ``cumulative'' measurements,
but in this work this is a satisfactory precision. Then we passed these data through a
periodogram computing tool of the PlanetPack software \citep{Baluev13c}, taking into
account the best fitting contribution from transiting planet (i.e., including it
in the null model of the periodogram).

In this analysis, none of the periodograms revealed hints of any additional variations
in the data. One exception is the GJ~436 case. In this case we have rather large amount of
the HARPS data, as well as Keck data from \citep{Knutson14}, and both these datasets
demonstrated clear hints of a red noise, similar to the one investigated e.g. by
\citet{Baluev13a}. The same might be true for HD189733, for which we also have large
amounts of the RV data. We plan to consider these cases detailedly in a separate work.

\section{The WASP-4 case}
\label{sec_WASP4}
\begin{figure*}
\includegraphics[width=0.99\textwidth]{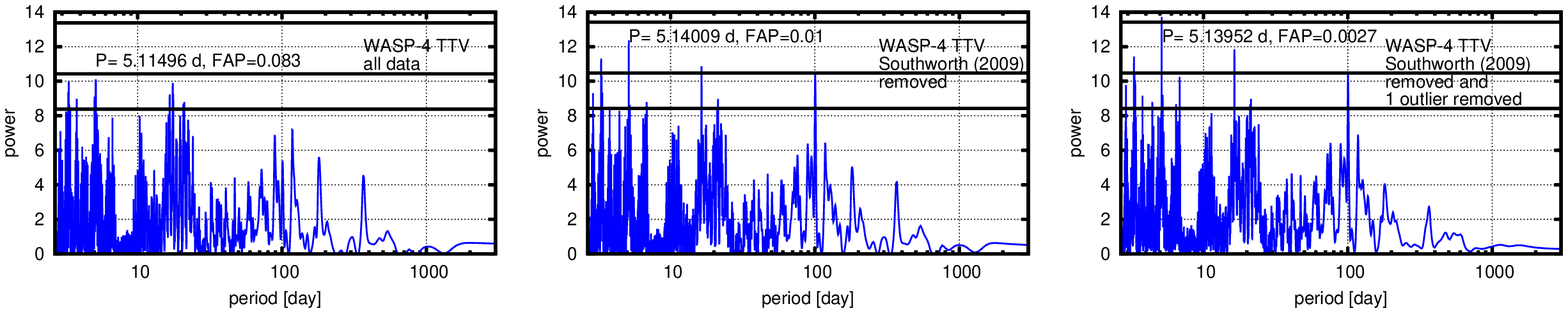}
\caption{Periodograms of the WASP-4 TTV data, showing a few marginally significant
peaks. The plot in the left panel was based on all available transit
lightcurves, the middle panel is for a reduced data set with the \citet{Southworth09b} data
removed, and in the right panel one TTV outlier detected in the \citep{Petrucci13} data was
also removed. The thick horizontal lines label the significance levels of $1\sigma$ or
$\FAP\approx 31.7$ per cent, $2\sigma$ or $\FAP\approx 4.6$ per cent, $3\sigma$ or
$\FAP\approx 0.27$ per cent. The significance and the period for the tallest
peak are printed in each panel. All periodograms involve a quadratic trend of the TTV delay
with fittable trend coefficients.}
\label{fig_WASP4pow}
\end{figure*}

The periodogram of the WASP-4 TTV data is shown in Fig.~\ref{fig_WASP4pow}, and it reveals
some marginally significant periodicities, corresponding to the TTV amplitudes $\sim
10-20$~sec. However, there is not any stable pattern of the peaks, as the periodograms are
severely model-dependent. Depending on which data subset we include in our analysis,
we obtain different results. Note that the data by \citet{Southworth09b} are not
reliable for TTV studies due to the clock failures noticed for this telescope. This is why
we also consider in Fig.~\ref{fig_WASP4pow} a reduced TTV time series obtained by removing
these data. There are only $4$ these TTV points, but they have a good formal accuracy as
derived from the photometric fit, and thus affect the TTV periodogram significantly. Also,
results of the period analysis depend on various other subtleties, e.g. on the degree of
the polynomial trend used in the photometry model. From the other side, although these
variations are marginal and model-dependent, we cannot just attribute them to the noise, as
we did not observe anything similar in the other stars of this study.

Using the method of Sect.~\ref{sec_ttv}, we detected an outlier in the
WASP-4 TTV data, owing to one of the \citet{Petrucci13} lightcurves. This peculiar
lightcurve was already noticed by \citet{Petrucci13} themselves. Interestingly, the height
of the periodogram peak at $\sim 5.14$~d is \emph{increased} after removal of this TTV
outlier from the analysis.

The RV data from the HARPS archive did not reveal any significant signal in addition to the
primary transiting planet. However, the star is rather faint for HARPS. Also, we should not
expect a direct connection between the TTV and RV periods, because often TTV is an
indirectly induced variation, owing to the planetary dynamical perturbations. Such a
variation may appear due to e.g. a mean motion resonance, which is not always easily
detected in radial velocities \citep[e.g.][]{Anglada-Escude09}.

Currently we remain uncertain about the nature of these variations that are possibly
present in the WASP-4 TTV data. Possibly, the TTV signal might be more complicated
than just a sinusoid, and it should be modelled in the framework of the Newtonian N-body
fits. Another possibility is star spots inducing systematic perturbations in the
timing measurements. In any case, it seems that we should keep tracking this target or
maybe even focus increased attention on it by making further observations in future.

\section{Conclusions and discussion}
As the main topic of the manuscript is the TTV exoplanet detection, it is interesting to
investigate its efficiency relatively to the more classic methods like the Doppler
planet detection. An apparent TTV signal can
be induced via two mechanisms: the dynamical perturbation on the transiter's motion and the
light arrival time delay due to the finite light speed (the Roemer
effect). The dynamical perturbations are difficult to predict in a general case, as they
severely depend on many orbital parameters (e.g. they may drastically increase when planet
move in a mean-motion resonance). But the Roemer delay can be easily assessed, so let us
now consider it detailedly.

Assume that a distant second planet has the mass of $m$ and orbits the
star (with the transiter) on a circular orbit with the radius of $a$. Such a planet should
induce a similar circular motion on the host star on an orbit of the radius $a' \simeq
m/M_\star a$, where $M_\star$ is the star mass. This reduces to a sinusoidal variation
of the transits time delay with an amplitude of
\begin{equation}
K_{\rm TTV} \simeq \frac{ma}{cM_\star} \sin i,
\label{Kttv}
\end{equation}
where $c$ is the speed of light, and $i$ is the orbital inclination to the sky plane.

In the same case, the amplitude of the Doppler variation induced on the star is equal to
\begin{equation}
K_{\rm RV} \simeq m\sin i \sqrt{\frac{G}{a M_\star}},
\label{Krv}
\end{equation}
where $G$ is the gravitational constant. The ratio of these amplitudes now looks like
\begin{equation}
\frac{K_{\rm TTV}}{K_{\rm RV}} \simeq \frac{a^{3/2}}{c\sqrt{G M_\star}} \simeq \frac{P}{2\pi c},
\end{equation}
where $P$ is the orbital period of the distant companion, computed according the the third
Kepler law. We can see that the relative efficiency of the TTV method, in comparison
with the RV one, only depends on the orbital period of the unseen companion.

Contrary to the RV and classic transit surveys, the TTV method might be useful to
detect \emph{distant} companions. Of course, in any case we must track the variation
over at least a single period, so detection of a distant companion necessarily
requires a long observation run. From this point of view, ground-based observatories should
be more useful for TTV planet detection. Spacecraft rarely operate over a term longer than
a few years, while ground-based observations can run on an indefinitely long time base.

The accuracy of the TTV data presented in this work is such that it would allow a robust
detection of a signal, if its amplitude is above $1$~min. With this TTV threshold, the
formula~(\ref{Kttv}) implies that a Jupiter-mass planet orbiting a Solar-mass star could be
detected, if its semimajor axis was at least $60$~AU, implying an extremely large period of
at least $\sim 400$~yr. This
is not a realistic requirement for the observation time. Alternatively, the planet
mass should be at least $12$ times the Jupiter mass, if we want to detect it
at a Jupiter-like orbit with $a=5.2$~AU. Also, the exoplanet detectability naturally
increases for small-mass hosts. Thus, to robustly detect a Jupiter-mass planet, we
should either wait for a long time or to decrease the TTV measurement errors to a level of
seconds, roughly by an order of magnitude. Based on these computations, we may say that the
TTV detection method is now in its early development stage, comparable to the early era of
the Doppler technique (prior or near the 51 Pegasi b detection).

Summarizing, we can conclude that amateur-class TTV observations may occupy two possible
niches: (i) detection of long-period massive exoplanets and (ii) detection of exoplanets
trapped in mean motion resonances with known transiters (thus generating a more remarkable
TTV signal via dynamical effects). However, the TTV detection of long-period
Jupiter-like planets might be a feasible task for more advanced ground-based observatories
that can achieve the TTV accuracy of a few seconds. Some of the data listed in
Table~\ref{tab_ref} do provide such an accuracy. To enable the detection of Jupiter twins,
such observations should be carried out in an experimental monitoring regime, i.e.
on a regular basis and over a long term, similar to the modern Doppler surveys. Note
that Jupiter analogues may represent a special interest, because
such planets would point out exoplanetary systems that have architectures similar to Solar
System, in which all giant planets are quite distant from the Sun. The chance to find an
Earth twin in such a system might be higher.

At last, we note that the transit fitting algorithm presented here is now implemented in
the free PlanetPack package \citep{Baluev13c} and is made available as
of the current version PlanetPack 2.1.

\section*{Acknowledgements}
This work was supported by the Russian Foundation for Basic Research
(project No. 14-02-92615 KO\_a), the UK Royal Society International Exchange grant
IE140055, by the President of Russia grant for young scientists (No. MK-733.2014.2), by the
programmes of the Presidium of Russian Academy of Sciences P21 and P22, by the Saint
Petersburg State University research grant 6.37.341.2015, and by the Russian Ministry of
Education and Science (contract No. 01201465056). OB acknowledges the support by the
research fund of Ankara University (BAP) through the project 13B4240006.
We express our gratitude to the anonymous reviewer for their useful suggestions.

\bibliographystyle{mn2e}
\bibliography{transits}

\appendix

\bsp

\label{lastpage}

\end{document}